\def\k{{\bm k}}
\def\x{{\bm x}}

\def\Im{\operatorname{Im}}

\def\zh{z_{\rm h}}
\def\rh{r_{\rm h}}
\def\Rads{L}

\def\idx{a}
\def\tort{r_*}
\def\torto{r_{*0}}
\def\m{\bar m}

\documentclass[prd,preprint,eqsecnum,nofootinbib,amsmath,amssymb,
               tightenlines,dvips]{revtex4}
\usepackage{graphicx}
\usepackage{bm}
\usepackage{amsfonts}
\usepackage{amssymb}
\usepackage{dsfont}

\begin {document}


\title
    {
      Spin \boldmath$1/2$ quasinormal mode frequencies in
      Schwarzschild-AdS spacetime
    }

\author{
  Peter Arnold and Phillip Szepietowski
}
\affiliation
    {%
    Department of Physics,
    University of Virginia, Box 400714,
    Charlottesville, Virginia 22904, USA
    }%

\date {\today}

\begin {abstract}%
{%
  We find the asymptotic formula for quasinormal mode frequencies $\omega_n$
  of the Dirac equation in a Schwarzschild-AdS$_D$ background in space-time
  dimension $D > 3$, in the large black-hole limit appropriate to many
  applications of the AdS/CFT correspondence.
  By asymptotic, we mean large overtone number $n$
  with everything else held fixed, and we find the $O(n^0)$ correction
  to the known leading $O(n)$ behavior of $\omega_n$.
  The result has the schematic form
  $\omega_n \simeq n \, \Delta\omega + A \, \ln n + B$,
  where $\Delta\omega$ and $A$ are constants and
  $B$ depends logarithmically on the $(D-2)$-dimensional
  spatial momentum $\k$ parallel to the horizon.
  We show that the asymptotic result agrees well with exact
  quasinormal mode frequencies computed numerically.
}%
\end {abstract}

\maketitle
\thispagestyle {empty}


\section {Introduction and Results}
\label{sec:intro}

\subsection {Overview}

There has been a great deal of numerical and analytic work on
quasinormal modes of black holes.  (For some modern reviews, see
Refs.\ \cite{QNMreview1,QNMreview2}.)  Because of
gauge-gravity duality, one case of
particular interest over the last decade has been that of black holes
in asymptotically anti-de Sitter spacetimes.
In this case, low-lying quasinormal mode frequencies can be related
to the poles of Green functions
\cite{Birmingham,SonStarinets,Starinets,KovtunStarinets} and so to
the long distance exponential fall-off of various correlators
in certain finite-temperature and finite-density field theories.
Also, the full set of quasinormal frequencies can be related
to functional determinants in the gravity theory,
which are dual to `$1/N$' corrections to strongly-coupled field
theories \cite{DHS,DHS0}.  Except in a few special cases, exact computation
of quasinormal modes requires numerics.  But it is generally possible,
using WKB-like methods, to obtain analytic results for quasinormal
mode frequencies $\omega_n$ in the limit of large overtone number $n$
--- that is, in the limit of large $|\omega_n|$.
In most cases, quasinormal mode frequencies are evenly spaced
at large $n$, with asymptotic expansion
\begin {equation}
   \omega_n \approx n \, \Delta\omega + \mbox{constant} + \cdots ,
\label {eq:expansion1}
\end {equation}
where $\Delta\omega$ is the asymptotically constant spacing
between successive $\omega_n$.  In most cases, the asymptotic
expansion has been computed analytically to at least $O(n^0)$,
so as to include the constant term on the right-hand side of
(\ref{eq:expansion1}).
In this paper, we focus on the limit where $n$ is taken large
while holding other quantities fixed (such as mass $m$ and
spatial momentum $\k$ parallel to the horizon).

The goal of the present paper is to fill one gap in the literature
concerning such asymptotic formulas for $\omega_n$.
To our knowledge, nobody has previously analyzed analytically the case of
quasinormal modes of half-integer spin fields in
Schwarzschild-AdS (SAdS) spacetimes through $O(n^0)$, except in the special
low-dimensional case of BTZ black holes \cite{BTZblackhole}
(in bulk space-time dimension $D{=}3$),
where exact analytic results for all $\omega_n$ may be found \cite{BTZdirac}.
In this paper, we specialize to the case of spin-$\tfrac12$ fields
in SAdS$_D$ with $D > 3$.  We will find a result
that has the schematic form
\begin {equation}
   \omega_n \approx n \, \Delta\omega + A \ln n + B + \cdots ,
\label {eq:expansion}
\end {equation}
and we will evaluate all the terms through $O(n^0)$.
An interesting and unusual feature here is the $\ln n$ term.
Such dependence on $\ln n$ does not arise for quasinormal modes
of asymptotically flat black holes
(see \cite{MusiriSiopsis:Dirac} for an analysis of the spin-$\tfrac12$ case),
nor does it arise in the analysis of
spin-0 or spin-2 quasinormal modes of SAdS
\cite{NatarioSchiappa,CardosoNatarioSchiappa,MusiriNessSiopsis}.%
\footnote{
   Depending on their background, some readers
   of refs.\ \cite{NatarioSchiappa,CardosoNatarioSchiappa}
   may need to be warned
   about terminology:
   the ``scalar-type,'' ``vector-type,'' and ``tensor-type''
   perturbations referred to there do not respectively
   refer to spin 0, 1, and 2
   fields but instead classify different components of gravitational
   perturbations.
   See, for example, the discussion
   of gravitational perturbations in section 2.1 of ref.\ \cite{QNMreview1}.
   However, the ``tensor-type'' case has the same equation of motion as
   a massless, minimally-coupled spin-0 field.
}
However, a logarithmic dependence has been found
for spin-1 quasinormal modes of SAdS for $D{=}4$ \cite{MusiriNessSiopsis}.
Our result shows that the spin-$\tfrac12$ case has a logarithm in
all dimensions $D>3$.
We will verify our result by comparison to
(i) existing numerical results for the special case of massless Dirac
fermions in $D{=}4$ \cite{GiammatteoJing}
and (ii) our own numeric calculations for
massive Dirac fermions in $D{=}5$.

We note in passing that the case of massive spin-$\tfrac12$ fields
in SAdS arises naturally in the duality between strongly-coupled
${\cal N}{=}4$ super Yang-Mills theory and
Type IIB supergravity on AdS$_5 \times S^5$ (or SAdS$_5 \times S^5$
for finite temperature) because the Kaluza-Klein reduction
on $S^5$ gives mass to all
spin-$\tfrac12$ fields in (S)AdS$_5$ \cite{KRvN}.

For the sake of simplicity of discussion and of AdS/CFT applications
to infinite-volume systems, we will focus on black branes, which
correspond to the limiting case of
arbitrarily ``large'' black holes.  In the remainder of this
introduction, we review the corresponding metric and set up our choice
of coordinates, and then we summarize our analytic results for the
asymptotic quasinormal mode frequencies.  The derivation is given in
section \ref{sec:derivation}.  In section \ref{sec:nmethod}, we
explain our numerical method for computing exact quasinormal mode
frequencies in the massive case.  Finally, comparison of our
asymptotic formulas to both our own numerics and the numerics of
ref.\ \cite{GiammatteoJing} is given in section \ref{sec:compare},
and the $\log(n)$ term in the expansion (\ref{eq:expansion})
is verified.

As mentioned earlier, in this paper we study the limit of large
overtone number $n$ with all other parameters held fixed.
Readers interested in other limits may find a discussion
of large chemical potential $\mu$ in
\cite{HerzogRen} (with specific calculations in a probe approximation).
Also, in the case of scalar rather than spin-$\tfrac12$ fields,
a discussion of
large mass $m$ may be found in refs.\ \cite{largem1,largem2}
and large momentum $\k$ in ref.\ \cite{largek}.


\subsection{Metric}

Two standard, equivalent forms of the SAdS$_D$ black-brane
metric in $D$ space-time
dimensions are
\begin {subequations}
\label {eq:metric}
\begin {equation}
   ds^2 = \frac{\Rads^2}{z^2} \left[ -f \, dt^2 + d\x^2 + f^{-1} dz^2 \right]
\end {equation}
and
\begin {equation}
   ds^2  = -F \, dt^2 + r^2 \, \frac{d\x^2}{\Rads^2} + F^{-1} dr^2 ,
\label {eq:metricr}
\end {equation}
\end {subequations}
where $\Rads$ is the radius associated with the asymptotic AdS space-time,
\begin {equation}
   z = \frac{\Rads^2}{r} \,,
\end {equation}
\begin {subequations}
\label {eq:fF}
\begin {equation}
   f
   \equiv 1 - \Bigl( \frac{\rh}{r} \Bigr)^{D-1}
   = 1 - \Bigl( \frac{z}{\zh} \Bigr)^{D-1} ,
\end {equation}
and%
\footnote{
   Many papers in the literature refer to our $F$ as $f$, in which case
   our $f$ would be their $f/r^2$ (in the large black hole limit).
}
\begin {equation}
   F \equiv \frac{r^2 f}{\Rads^2} = \frac{\Rads^2 f}{z^2} \,.
\end {equation}
\end {subequations}
The boundary of AdS is at $z=0$ and $r=\infty$, the black hole horizon
is at $z=\zh$ and $r=\rh$, and the singularity is at $z=\infty$
and $r=0$.
(Since both $z$ and $r$ are common variable choices, depending on context,
we will occasionally
jump back and forth between them in order to facilitate comparison of
formulas with the rest of the literature on quasinormal modes.)

We will consider fermions of mass $m$ propagating in this metric background.
In applications of gauge-gravity duality, masses $m$ of the spin-$\tfrac12$ fields in the gravity theory
are related by duality
to conformal dimensions $\Delta$ of spin-$\tfrac12$ operators in the field
theory by $|m\Rads| = \Delta - \frac12(D-1)$ \cite{MAGOO}.


\subsection{Asymptotic Results}

The result we find in this paper for (retarded) quasinormal mode
frequencies of massive spin $\tfrac12$ fermions in SAdS is
\begin {multline}
   \frac{\omega_{n\pm}}{\pi T} \simeq
   4 e^{-i \pi/(D-1)} \sin\Bigl( \frac{\pi}{D-1} \Bigr)
\\ \times
   \left[
      n
      - \frac{i}{2\pi} \ln\left( \frac{n^\idx}{|\k|/\pi T} \right)
      + \tfrac12(|m\Rads|-1\pm\tfrac12)
      + \frac{(D-3)^2}{8(D-1)(D-2)}
      + \frac{i \xi}{\pi}
   \right]
\label {eq:fermion}
\end {multline}
\begin {equation*}
   \mbox{(massive Dirac fermion, $D>3$),}
\end {equation*}
where
$\k$ is the $(D{-}2)$-dimensional spatial momentum
conjugate to $\x$ in (\ref{eq:metric}),
\begin {equation}
   T = \frac{(D-1)\rh}{4\pi\Rads^2} = \frac{(D-1)}{4\pi\zh}
\label {eq:T}
\end {equation}
is the Hawking temperature of the black hole,
\begin {equation}
   \xi \equiv
   - \frac{\idx}{2} \ln \left( 4 \sin\Bigl(\frac{\pi}{D-1}\Bigr) \right)
   + \frac{(1-\idx)}{2} \ln(1-\idx)
   + \frac12 \ln\bigl[ \Gamma(\idx) \sin(\pi\idx) \bigr] ,
\label {eq:xi}
\end {equation}
and
\begin {equation}
   \idx \equiv \frac{(D-3)}{2(D-2)} \,.
\label {eq:idx}
\end {equation}
We assume $\k\not=0$ here and throughout.
The leading $O(n)$ piece of our result (\ref{eq:fermion})
was known previously from numerical results on the asymptotic spacing
between modes for massless spin-$\frac12$
fermions in $D{=}4$ \cite{GiammatteoJing} and
is the same for general $D$
as the leading $O(n)$ result for fields of other
spin \cite{MusiriSiopsis:AdS,MusiriNessSiopsis,NatarioSchiappa}.%
\footnote{
  This is not an accident, as the monodromy arguments used to determine
  asymptotic behavior analytically do not depend on spin at leading
  order in $n$.
}
Some authors adopt the convention of studying advanced rather than
retarded quasinormal modes (outgoing rather than infalling boundary
condition at the horizon), which corresponds to taking the
complex conjugate of our result.

The $\pm$ sign in (\ref{eq:fermion}) distinguishes
different spin states, in a way to be made precise later.
The formula (\ref{eq:fermion}) only shows the quasinormal mode
frequencies $\omega = \omega_{n\pm}$
in the right-half complex plane.  The corresponding quasinormal
frequencies in the left-hand plane are given by
$\omega = -\omega_{n\mp}^*$.

\begin {figure}
\begin {center}
  \includegraphics[scale=0.75]{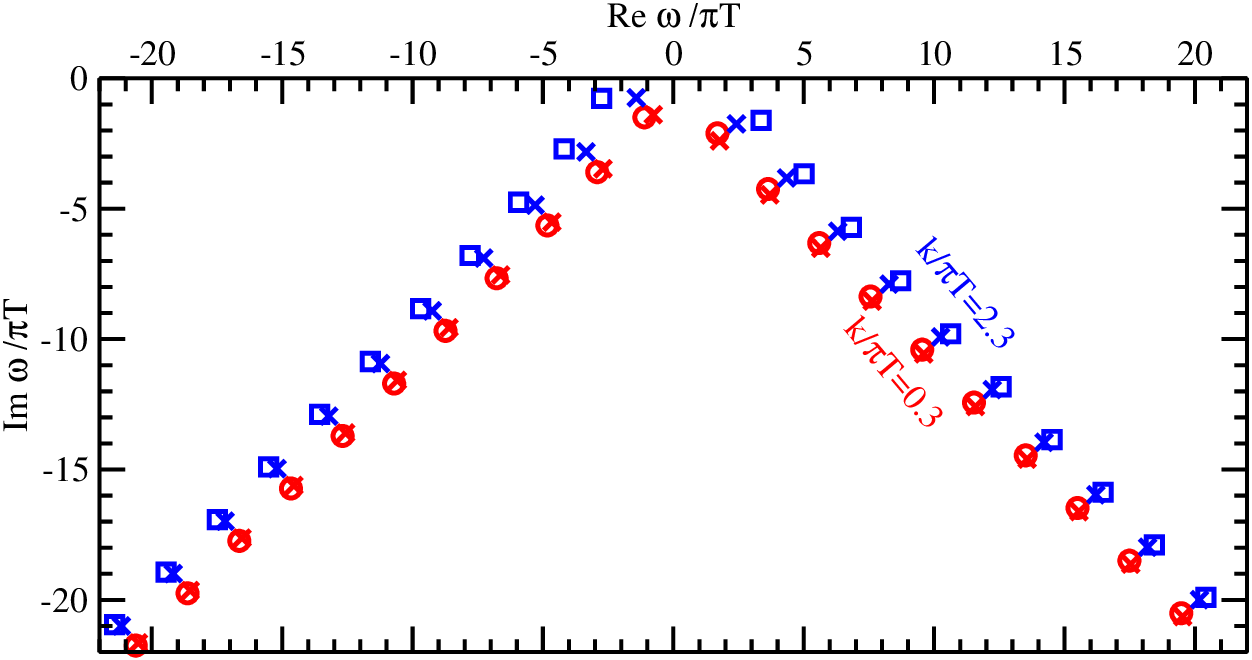}
  \caption{
     \label{fig:wplane}
     Retarded quasi-normal mode frequencies in the
     complex $\omega$ plane for $D{=}5$, $m\Rads=\tfrac12$,
     and spin states with
     ${\bm\sigma}\cdot{\bm\hat\k} = +1$ (as defined in section
     \ref{sec:dirac}).
     Note that the figure is not left-right symmetric.
     The result for ${\bm\sigma}\cdot{\bm\hat\k} = -1$ is the
     left-right mirror image of this figure, as required by parity.
     The red circles are numerical results
     for $|\k|/\pi T = 0.3$, and the blue squares are those for
     $|\k|/\pi T = 2.3$.  The correspondingly colored crosses show
     the asymptotic formula (\ref{eq:D5}).
  }
\end {center}
\end {figure}

For $D{=}5$, our result specializes to
\begin {equation}
   \frac{\omega_{n\pm}}{\pi T} \simeq
   2(1-i)
   \left[
      n
      - \frac{i}{2\pi} \ln\left( \frac{n^{1/3}}{|\k|/\pi T} \right)
      + \tfrac12(|m\Rads|-1\pm\tfrac12)
      + \frac{1}{24}
      + \frac{i \xi_5}{\pi}
   \right]
\label {eq:D5}
\end {equation}
\begin {equation*}
   \mbox{(massive Dirac fermion, $D=5$),}
\end {equation*}
with
\begin {equation}
   \xi_5 =
   \tfrac12 \ln \Gamma(\tfrac13) - \tfrac1{12} \ln3 - \tfrac{5}{12}\ln2
   = 0.112348 .
\end {equation}
Henceforth, we will assume that $m$ is chosen by
convention to be positive, and so we will drop the absolute value signs
in (\ref{eq:fermion}) and elsewhere.
Fig.\ \ref{fig:wplane} gives a first look at the comparison of our
numerical results (described later)
for exact quasi-normal mode frequencies
and the asymptotic formula (\ref{eq:D5}) for $D{=}5$,
$m\Rads = \tfrac12$, and two different values of
$|\k|$.  We will make more precise comparisons later on.

{\it Massless}\/ fermions are a special case and require a careful
discussion of boundary conditions, to be given later.
Massless fermions have previously been studied numerically
by Giammatteo and Jing \cite{GiammatteoJing} in $D{=}4$.
We will find that, for their choice of boundary conditions,
the analytic result for large $n$ happens
to be given by the same mathematical formula
as (\ref{eq:fermion}) if you plug in $|m\Rads|=\tfrac32$ instead
of the naive choice $|m\Rads|=0$.  For comparison to the
$D{=}4$ simulations of Giammatteo and Jing, whose numerical results
are for the case we call $\omega_{n+}$ in the lower-right complex plane,
our result is then
\begin {equation}
   \frac{\omega_{n+}}{\pi T} \simeq
   e^{-i \pi/3} 2\sqrt3
   \left[
      n
      - \frac{i}{2\pi} \ln\left( \frac{n^{1/4}}{|\k|/\pi T} \right)
      + \frac{25}{48}
      + \frac{i \xi_4}{\pi}
   \right]
\label {eq:fmassless4}
\end {equation}
\begin {equation*}
   \mbox{(massless Dirac fermion, $D=4$),}
\end {equation*}
with
\begin {equation}
   \xi_4 =
   \tfrac12 \ln \Gamma(\tfrac14) + \tfrac5{16} \ln3 - \tfrac98 \ln2
   = 0.207537 \,.
\end {equation}


\section{Derivation}
\label {sec:derivation}

\subsection{The Dirac equation}
\label {sec:dirac}

The Dirac equation in curved space is
\begin {equation}
   \Gamma^M D_M \Psi - m \Psi = 0
\end {equation}
where $\{\Gamma^M,\Gamma^N\}=2g^{MN}$ and where the covariant derivative
$D_M$ contains a spin-connection term.
As noted by Herzog and Ren \cite{HerzogRen}, the Dirac equation may be
simplified in the case of metrics of the form
\begin {equation}
   ds^2 = g_{tt}(z) \, dt^2 + g_{xx}(z) \, d\x^2 + g_{zz}(z) \, dz^2
\end {equation}
by rescaling
\begin {equation}
  \psi \equiv (-g g^{zz})^{1/4} \Psi .
\end {equation}
The result is
\begin {equation}
   \left[
     \sqrt{-g^{tt}} \, \gamma^t \partial_t
     + \sqrt{g^{xx}} \, \gamma^i \partial_i
     + \sqrt{g^{zz}} \, \gamma^z \partial_z
     - m
   \right] \psi = 0,
\end {equation}
where the $\gamma^m$ are flat-spacetime $\gamma$ matrices with
$\{\gamma^m,\gamma^n\}=2\eta^{mn}$
and the index $i$ runs over the $D{-}2$ dimensions of $\x$.
Equivalently, working in momentum space $k_\mu = (-\omega,\k)$ for
all coordinates except $z$,
\begin {equation}
   \left[
     - i \sqrt{-g^{tt}} \, \gamma^t \omega
     + i \sqrt{g^{xx}} \, \gamma^i k_i
     + \sqrt{g^{zz}} \, \gamma^z \partial_z
     - m
   \right] \psi = 0,
\end {equation}
In the case of the SAdS metric (\ref{eq:metric}), this is
\begin {equation}
  \psi \equiv (\Rads z)^{-(D-1)/2} f^{1/4} \Psi
\end {equation}
giving
\begin {equation}
  \left[
     - i z f^{-1/2} \gamma^t \omega
     + i z {\bm\gamma}\cdot\k
     + z f^{1/2} \gamma^z \partial_z
     - m \Rads
   \right] \psi = 0 .
\label {eq:Dirac1}
\end {equation}
It is standard and convenient to split $\psi$ into two pieces
$\psi_\pm$ according to their chirality under $\gamma^z$.
We choose a representation of the Dirac matrices of the
form
\begin {equation}
  \gamma^{z} =
    \begin{pmatrix} \openone & \\ & -\openone \end{pmatrix}
    = \tau_3 \otimes \openone ,
  \qquad
  \gamma^t =
    i \begin{pmatrix} 0 & \openone \\ \openone & 0 \end{pmatrix}
    = i \tau_1 \otimes \openone ,
  \qquad
  {\bm\gamma} =
    \begin{pmatrix} 0 & i{\bm\sigma} \\ -i{\bm\sigma} & 0 \end{pmatrix}
    = -\tau_2 \otimes {\bm\sigma} .
\end {equation}
Here the $\tau_i$ are Pauli matrices that mix the $\psi_+$ and $\psi_-$
of
\begin {equation}
   \psi = \begin{pmatrix} \psi_+ \\ \psi_- \end{pmatrix} ,
\end {equation}
and the ${\bm\sigma}$ are $D{-}2$ anti-commuting matrices with
$\{\sigma_i,\sigma_j\}=2\delta_{ij}$ (also Pauli matrices in
the cases $D{=}4$ and $D{=}5$).  In this notation, the Dirac equation
(\ref{eq:Dirac1}) becomes (multiplying by $f^{1/2} \gamma^z/z$)
\begin {equation}
   \left[-f \partial_z  - i \omega \tau_2 + m \Rads \, \frac{f^{1/2}}{z} \, \tau_3
      + f^{1/2} \tau_1 {\bm\sigma}\cdot\k \right] 
   \begin{pmatrix} \psi_+ \\ \psi_- \end{pmatrix}
   = 0
\end {equation}
or equivalently%
\footnote{
  An equivalent equation may be found in ref.\ \cite{Giecold} in
  terms of the original $\Psi_\pm$ rather than $\psi_\pm$.
  There is a sign convention difference in the subscript of $\Psi_\pm$
  related to whether one defines the sign by $\gamma^z$ or $\gamma^r$.
}
\begin {equation}
   \left[F \partial_r - i \omega \tau_2 + m F^{1/2} \tau_3
      + f^{1/2} \tau_1 {\bm\sigma}\cdot\k \right] 
   \begin{pmatrix} \psi_+ \\ \psi_- \end{pmatrix}
   = 0 .
\end {equation}
The combination $-f\partial_z = F\partial_r$ is just the derivative
$\partial_{\tort}$ with respect to the tortoise coordinate $\tort$ defined
by
\begin {equation}
   d\tort \equiv - \frac{dz}{f} = \frac{dr}{F} \,.
\label {eq:tortdef}
\end {equation}
We may choose the solutions to be eigenstates of ${\bm\sigma}\cdot\k$
(or equivalently $\gamma^z \gamma^t {\bm\gamma}\cdot\k$)
with eigenvalue
\begin {equation}
   k \equiv \pm |\k| .
\label {eq:k}
\end {equation}
The $\pm$ sign above is the same $\pm$ used to distinguish the different
cases in our final result (\ref{eq:fermion}).
With this notation,
\begin {equation}
   \left[\partial_{\tort} - i \omega \tau_2 + m F^{1/2} \tau_3
      + k f^{1/2} \tau_1 \right]
   \begin{pmatrix} \psi_+ \\ \psi_- \end{pmatrix}
   = 0 .
\label {eq:Dirac}
\end {equation}

The quasinormal mode boundary conditions that we
will apply are that (i) $\psi$ vanishes at the boundary of AdS
(except in the massless case to be discussed later) and (ii) $\psi$
is infalling at the horizon, i.e. $\psi \sim e^{-i\omega\tort}$ near
the horizon.
Note that if $\psi=\tilde\psi(\omega_n,K,r)$ is a solution to
(\ref{eq:Dirac}) with these boundary conditions for
complex $\omega = \omega_n$ and real $k = K$, then
$\psi = \tau_3[\tilde\psi(\omega_n,K,r)]^*$
also solves (\ref{eq:Dirac}) and satisfies
the boundary conditions, but with $\omega = -\omega_n^*$
and $k = -K$.
This transformation maps solutions in the right-half
complex $\omega$ plane with one sign $\pm$ of ${\bm\sigma}\cdot{\bm\hat\k}$
to solutions in the left-half complex plane with the other sign $\mp$
of ${\bm\sigma}\cdot{\bm\hat\k}$.
In the discussion that follows, we will focus on the solutions in
the right-half plane.


\subsection {The Method}

\begin {figure}
\begin {center}
  \includegraphics[scale=0.75]{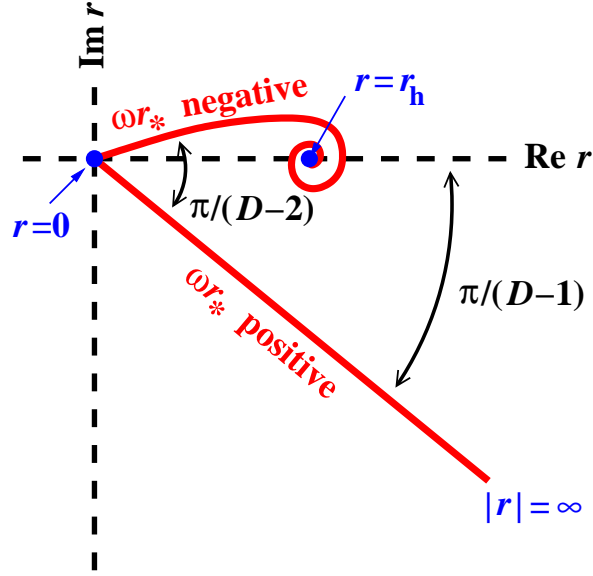}
  \caption{
     \label{fig:stokes}
     A qualitative picture of the relevant Stokes lines $\Im(\omega\tort)=0$
     in the complex $r$ plane
     for following WKB between the boundary
     ($r=\infty$) and the horizon ($r=\rh$).
     The path passes through the singularity ($r=0$) in SAdS$_D$.
     The other Stokes lines emanating from the origin are not shown, one
     of which escapes to $-e^{-i\pi/(D-1)}\infty$ and the others which
     spiral into the complex-valued horizons $r = e^{-i2n\pi/(D-1)}\rh$
     for $n=1,\cdots,D-2$.
     (The spiral into the horizon shown above crosses a cut in the
     definition of $\tort$ emanating
     from $r=\rh$, and the curve spirals onto higher and higher
     Riemann sheets.)  Given our retarded convention for $\omega$,
     this figure is the complex conjugate of similar diagrams in
     refs.\ \cite{NatarioSchiappa,CardosoNatarioSchiappa}.
  }
\end {center}
\end {figure}

To find the asymptotic quasinormal mode frequencies, we will use the
Stokes line method nicely reviewed in ref.\ \cite{NatarioSchiappa}.
Start by taking the naive large-$\omega$ limit of (\ref{eq:Dirac}),
which is
\begin {equation}
   \left[\partial_{\tort} - i \omega \tau_2 \right]
   \begin{pmatrix} \psi_+ \\ \psi_- \end{pmatrix}
   \approx 0
\label {eq:naive}
\end {equation}
and has solution
\begin {equation}
   \psi \simeq A_+ e^{i\omega\tort} + A_- e^{-i\omega\tort} .
\label {eq:WKBnaive}
\end {equation}
One difficulty with this approximation is that there are regions
where the other terms in (\ref{eq:Dirac}) may not be ignored, even
when $\omega$ is large.  This happens near the boundary ($r=\infty$)
and the singularity ($r=0$), and so one must do a matching calculation
if one wishes to follow the solution there.  A more serious difficulty
is that the quasinormal frequencies $\omega$ that we are looking for
have imaginary parts, and so one of the two terms
in the solution (\ref{eq:WKBnaive}) will become exponentially small
compared to the other as we follow the solution from the boundary to
the horizon, and so that term
becomes smaller than the error of the large-$\omega$
solution.
To avoid this difficulty, one lifts $r$ from the real axis to the complex
plane and traces Stokes lines, defined by $\Im(\omega\tort)$ = 0,
for which the magnitude of the two terms in (\ref{eq:WKBnaive})
remain the same size.  For analyzing asymptotic quasinormal mode
frequencies, it is adequate to use the leading-order $O(n)$
formula for $\omega$ in the Stokes condition, which in SAdS has complex
phase $\omega \propto e^{-i\pi/(D-1)}$ as in (\ref{eq:fermion}).
So the Stokes lines are given by
\begin {equation}
   \Im \left( e^{-i\pi/(D-1)} \tort \right) \simeq 0 .
\label {eq:stokes}
\end {equation}
We choose the tortoise coordinate $\tort$
to be zero at the singularity $r=0$, in which case
(\ref{eq:tortdef}) gives
\begin {equation}
   \tort =
   \sum_{n=1}^{D-1}
       \frac{\ln\bigl(1-\frac{r}{r_n}\bigr)}{F'(r_n)}
   =
   \frac{\Rads^2}{(D-1)\rh} \sum_{n=1}^{D-1}
         e^{i 2n\pi/(D-1)} \ln\left(1 - \frac{r}{\rh}\, e^{i 2n\pi/(D-1)}\right) ,
\label {eq:tort}
\end {equation}
where $r_n = e^{-i 2n\pi/(D-1)} \rh$ are the roots of $F(r)$ and we have
written the formula in a way appropriate for our choice of cuts in
later discussion.  Following the path of discussion in
ref.\ \cite{NatarioSchiappa,CardosoNatarioSchiappa}
(which so far is independent of the spin of the field),
a qualitative sketch of the particular Stokes lines (\ref{eq:stokes}) that we
will use is given in fig.\ \ref{fig:stokes}.
By following these Stokes lines, we can relate the boundary condition
at $r{=}\infty$ to the solution near the singularity $r{=}0$ and thence in
turn to the boundary condition at the horizon $r{=}\rh$.
The WKB solution (\ref{eq:WKBnaive}) is not valid
very close to the boundary or to the singularity, so we will have
to separately solve the Dirac equation in those limiting cases in order
to match to the WKB solutions.

Given our conventions, the tortoise coordinate $\tort$ vanishes
at the singularity $r{=}0$, the Stokes line from the singularity to
$|r|{=}\infty$ has $\omega\tort$ positive, and the Stokes line from
the singularity to the horizon $r{=}\rh$ has $\omega\tort$ negative.

Note, by the way, that (\ref{eq:WKBnaive}) is the form of the
{\it exact}\/ solution
to the Dirac equation (\ref{eq:Dirac}) in the case that both $m$ and
$\k$ are zero.  However, (\ref{eq:WKBnaive}) cannot simultaneously
satisfy the quasinormal mode boundary conditions that
it vanish at the boundary of
AdS and that it have only the $e^{-i\omega\tort}$ component at the
horizon.  The existence of any quasinormal mode solution therefore
depends on
non-zero $m$ or $\k$.  This will be the origin of why
our asymptotic formula (\ref{eq:fermion}) for $\omega_n$
depends on $|\k|$.


\subsection {Matching at the boundary \boldmath$r \to \infty$}

Near the boundary, $f \to 1$ and the SAdS metric approaches that
of pure AdS.  The Dirac equation (\ref{eq:Dirac}) reduces to
\begin {equation}
   \left[\frac{r^2}{\Rads^2} \, \partial_r - i \omega \tau_2
      + \frac{m r}{\Rads} \, \tau_3
      + k \tau_1 \right]
   \begin{pmatrix} \psi_+ \\ \psi_- \end{pmatrix}
   \simeq 0
\end {equation}
or equivalently
\begin {equation}
   \left[\partial_z + i \omega \tau_2 - \frac{m\Rads}{z} \, \tau_3
      - k \tau_1 \right]
   \begin{pmatrix} \psi_+ \\ \psi_- \end{pmatrix}
   \simeq 0 ,
\label {eq:ads1}
\end {equation}
whose solution is well known \cite{Henningson,Leigh}.
Applying $-\partial_z + i \omega \tau_2 - m \Rads z^{-1} \tau_3 - k \tau_1$
to (\ref{eq:ads1}),
\begin {equation}
   \left[-\partial_z^2 + \frac{(2m\Rads-1)^2-1}{4 z^2}
      \right] \psi_+
   =
   \Omega^2 \psi_+ ,
\end {equation}
\begin {equation}
   \left[-\partial_z^2 + \frac{(2m\Rads+1)^2-1}{4 z^2}
      \right] \psi_-
   =
   \Omega^2 \psi_- ,
\end {equation}
where
\begin {equation}
   \Omega^2 \equiv \omega^2 - |\k|^2 .
\end {equation}
Independent solutions may be expressed in terms of Bessel functions
as $\sqrt{\Omega z} \, J_\nu(\Omega z)$ and
$\sqrt{\Omega z} \, Y_\nu(\Omega z)$ with $\nu = m\Rads \mp \tfrac12$
for $\psi_\pm$.
Note that solutions for
$\psi_+$ and $\psi_-$ are not independent from each other but are related
by the original equation (\ref{eq:ads1}).


\subsubsection {Massive fields}

For massive fields, we may impose the boundary
condition that the field $\psi$ vanish at the boundary $z{=}0$, which selects
the $J_\nu$ solutions.  The corresponding solution to the
original first-order equation (\ref{eq:ads1}) is
\begin {equation}
   \begin{pmatrix} \psi_+ \\ \psi_- \end{pmatrix}
   \propto
   \begin{pmatrix}
     \sqrt{2\pi\Omega z} \, J_{m\Rads-\frac12}(\Omega z) \\
     \frac{\Omega}{\omega-k} \,
        \sqrt{2\pi\Omega z} \, J_{m\Rads+\frac12}(\Omega z)
   \end{pmatrix}
   .
\label {eq:boundary1}
\end {equation}
We're interested in the large-$\omega$ limit of $|\omega| \gg |\k|$,
in which case this becomes
\begin {equation}
   \begin{pmatrix} \psi_+ \\ \psi_- \end{pmatrix}
   \propto
   \begin{pmatrix}
     \sqrt{2\pi\omega z} \, J_{m\Rads-\frac12}(\omega z) \\
     \sqrt{2\pi\omega z} \, J_{m\Rads+\frac12}(\omega z)
   \end{pmatrix}
\end {equation}
with asymptotic expansion
\begin {equation}
   \begin{pmatrix} \psi_+ \\ \psi_- \end{pmatrix}
   \propto
   \begin {pmatrix}
     2 \cos(\omega z - \tfrac{\pi}{2}\,m\Rads) \\
     2 \sin(\omega z - \tfrac{\pi}{2}\,m\Rads)
   \end{pmatrix}
   =
   \begin {pmatrix}
     e^{-im\Rads\pi/2}e^{i\omega z} + e^{im\Rads\pi/2}e^{-i\omega z} \\
     -i e^{-im\Rads\pi/2}e^{i\omega z} + i e^{im\Rads\pi/2}e^{-i\omega z}
   \end {pmatrix}
\end {equation}
away from the boundary (i.e.\ for $\omega z \gg 1$).
It will be useful later to rewrite this in terms of $\tau_1$ as
\begin {equation}
   \begin{pmatrix} \psi_+ \\ \psi_- \end{pmatrix}
   \propto
   \left[ -i \tau_1 e^{-im\Rads\pi/2} e^{i\omega z}
          + e^{im\Rads\pi/2} e^{-i\omega z} \right]
   \begin{pmatrix} 1 \\ i \end{pmatrix}
   .
\label {eq:WKBboundary}
\end {equation}

In order to move beyond the $z{\ll}\zh$ approximation, we need to
match the asymptotic form (\ref{eq:WKBboundary}) to the general
WKB form (\ref{eq:WKBnaive}).  To that end, we need the
near-boundary expression for the tortoise coordinate $\tort$
of (\ref{eq:tort}), which is
\begin {equation}
   \tort \simeq \torto - z ,
\end {equation}
where, for the $f$ specific to SAdS (\ref{eq:fF}),
\begin {equation}
   \torto \equiv \tort \bigr|_{\rm boundary} =
   \frac{\pi e^{i \pi/(D-1)}}{F'(\rh) \sin(\frac{\pi}{D-1})}
   = \frac{e^{i \pi/(D-1)}}{4 T \sin(\frac{\pi}{D-1})} .
\label {eq:x0}
\end {equation}
And so (\ref{eq:WKBboundary}) may be written as
\begin {equation}
   \psi
   \propto
   \left[ e^{im\Rads\pi/2} e^{-i\omega\torto} e^{i\omega\tort}
          -i \tau_1 e^{-im\Rads\pi/2} e^{i\omega\torto} e^{-i\omega\tort} \right]
   \begin{pmatrix} 1 \\ i \end{pmatrix}
   .
\label {eq:boundarymatch}
\end {equation}
It will be useful later to imagine expanding the result in
terms of $\tau_1$ eigenstates.
We won't actually need to be any more explicit than we already have been, but
one could accordingly rewrite
\begin{equation}
   \begin{pmatrix} 1 \\ i \end{pmatrix}
   \propto
   e^{i\pi/4}
   \begin{pmatrix} 1 \\ 1 \end{pmatrix}
   + e^{-i\pi/4}
   \begin{pmatrix} 1 \\ -1 \end{pmatrix}
\end {equation}
above, if desired.


\subsubsection {Massless fields}

For the massless case $m{=}0$, the solutions involving $J_\nu$ and
$Y_\nu$ near the boundary are
\begin {equation}
   \begin{pmatrix} \psi_+ \\ \psi_- \end{pmatrix}
   \propto
   \begin{pmatrix}
      \cos(\Omega z) \\
      \frac{\Omega \sin(\Omega z)}{\omega-k}
   \end {pmatrix}
\end {equation}
and
\begin {equation}
   \begin{pmatrix} \psi_+ \\ \psi_- \end{pmatrix}
   \propto
   \begin{pmatrix}
      \sin(\Omega z) \\
      - \frac{\Omega \cos(\Omega z)}{\omega-k}
   \end {pmatrix}
\end {equation}
respectively, and no non-trivial combination vanishes at the boundary.
Our interest in this case will be for comparison with the numerical
results of Giammatteo and Jing \cite{GiammatteoJing},
and so we should apply their boundary
conditions, which in our language is that $\psi_+=\psi_-$ at
the boundary.%
\footnote{
   Eq.\ (2.18) of Giammatteo and Jing \cite{GiammatteoJing} is the same
   as the massless version of our (\ref{eq:Dirac}) if one identifies
   their $F$ and $G$ with our $\psi_- - \psi_+$ and $\psi_- + \psi_+$
   respectively, their $f$ with our $F = r^2 f/\Rads^2$, and their $k_\pm$ with
   our $k\Rads$.  They subsequently impose the boundary condition that
   their $F$ vanish at the boundary of AdS.
}
In the large $\omega$ limit of interest, this fixes the combination
to be
\begin {equation}
   \begin{pmatrix} \psi_+ \\ \psi_- \end{pmatrix}
   \propto
   \begin{pmatrix}
      \cos(\omega z) - \sin(\omega z) \\
      \sin(\omega z) + \cos(\omega z)
   \end {pmatrix} ,
\end {equation}
which can be rewritten as
\begin {equation}
   \begin{pmatrix} \psi_+ \\ \psi_- \end{pmatrix}
   \propto
   e^{-i\pi/4}
   \left[ \tau_1 e^{i\omega z}
          + e^{-i\omega z} \right]
   \begin{pmatrix} 1 \\ i \end{pmatrix}
   \simeq
   e^{-i\pi/4}
   \left[ e^{-i\omega\torto} e^{i\omega\tort}
          + \tau_1 e^{i\omega\torto} e^{-i\omega\tort} \right]
   \begin{pmatrix} 1 \\ i \end{pmatrix} .
\end {equation}
This $m{=}0$ case with Giammatteo's and Jing's boundary condition
has the same asymptotic expansion as the $m\Rads=\tfrac32$ case of
(\ref{eq:boundarymatch}), which implemented the usual boundary condition
of avoiding singularities at the boundary for $m > 0$.
As it turns out, the matching near the boundary is the only place where
the mass $m$ will enter the calculation of the quasi-normal mode
frequencies to the order at which we are working.  So we need not
discuss the massless case any further: to compare to
Giammatteo and Jing, just set $D=4$ and $m\Rads=\tfrac32$ in the
formulas that we will derive for the massive case.


\subsection {Matching at \boldmath$r\to0$}

From the previous discussion, we now know the asymptotic expansion
along the Stokes line from $|r|=\infty$ to the singularity $r=0$
in fig.\ \ref{fig:stokes}.
For now, let's generically refer to this
expansion as
\begin {equation}
   \phi \simeq {\cal B}_+ e^{i\omega\tort} + {\cal B}_- e^{-i\omega\tort}
   \qquad \mbox{($\omega\tort$ positive)} ,
\label {eq:fWKBb}
\end {equation}
and we will save for later using the explicit from (\ref{eq:boundarymatch}).
Our task now is to solve the Dirac equation near
the singularity and thereby relate the coefficients ${\cal B}_\pm$
above to coefficients ${\cal A}_\pm$
for a similar asymptotic expansion along the other
Stokes line in fig.\ \ref{fig:stokes}, which leads to the horizon:
\begin {equation}
   \phi \simeq {\cal A}_+ e^{i\omega\tort} + {\cal A}_- e^{-i\omega\tort}
   \qquad \mbox{($\omega\tort$ negative)}.
\label {eq:fWKBa}
\end {equation}


\subsubsection{The $r\to0$ Schr\"odinger problem}

For small $r$ and large $\omega$, the dominant terms in the Dirac
equation (\ref{eq:Dirac}) are%
\footnote{
  We have chosen the sign of the $-i k$ term in (\ref{eq:sing1}) by
  going around the singularity at $r=\rh$ in the lower-half complex
  $r$ plane, according to the Stokes line from the boundary to the
  singularity in fig.\ \ref{fig:stokes}.
}
\begin {equation}
   \left[F \partial_r - i \omega \tau_2
      - i k \left(\frac{\rh}{r}\right)^{(D-1)/2} \tau_1 \right]
   \begin{pmatrix} \psi_+ \\ \psi_- \end{pmatrix}
   \simeq 0 ,
\label {eq:sing1}
\end {equation}
assuming $\k\not=0$.
Now apply
$-F \partial_r - i \omega \tau_2 - i k (\rh/r)^{(D-1)/2} \tau_1$ to
(\ref{eq:sing1}), and expand.
The result is a
Schr\"odinger-like equation
\begin {equation}
   \left[-\partial_{\tort}^2 + V_{\rm sing}(r)\right] \psi \simeq \omega^2\psi ,
\end {equation}
with
\begin {equation}
   V_{\rm sing}(r) =
   - k^2 \left( \frac{\rh}{r} \right)^{D-1}
   + i \tau_1 k \, \frac{(D-1)\rh^{3(D-1)/2}}{2\Rads^2\,r^{(3D-5)/2}}
   .
\label {eq:Vsing0}
\end {equation}
For the case of $D>3$ considered in this paper,
the first term in (\ref{eq:Vsingf1}) is sub-dominant for
small $r$ and so may be dropped:
\begin {equation}
   V_{\rm sing}(r) \simeq
   + i \tau_1 k \, \frac{(D-1)\rh^{3(D-1)/2}}{2\Rads^2\,r^{(3D-5)/2}}
   .
\label {eq:Vsingf1}
\end {equation}

Now rewrite this potential in terms of the tortoise coordinate $\tort$.
For small $r$, (\ref{eq:tortdef}) and/or (\ref{eq:tort}) give
\begin {equation}
  \tort = \int_0^r \frac{dr'}{F(r')} \simeq
  - \frac{\Rads^2 r^{D-2}}{(D-2) \rh^{D-1}} ,
\end {equation}
and so%
\footnote{
   The branch in (\ref{eq:rvsx2})
   has been chosen so that $\omega\tort$ positive real with
   $\arg\omega = -\arg\tort = -i\pi/(D-1)$
   corresponds to the Stokes line
   between the singularity and boundary shown in fig.\ \ref{fig:stokes}.
}
\begin {equation}
  r \simeq
  e^{-i\pi/(D-2)} \left[ \frac{(D-2)\rh^{D-1}\tort}{\Rads^2} \right]^{1/(D-2)} \,.
\label {eq:rvsx2}
\end {equation}
Then
\begin {equation}
   V_{\rm sing} \simeq - \tau_1 \frac{\kappa}{\tort^{2-\idx}}
\end {equation}
with $\idx$ defined as in (\ref{eq:idx})
and
\begin {equation}
   \kappa \equiv
   e^{i\pi(\frac32-\idx)} \, \frac{(D-1)}{2(D-2)^{2-\idx}}
   \left(\frac{\Rads^2}{\rh}\right)^{1-\idx} k .
\label {eq:kappa}
\end {equation}


\subsubsection{The solution}
\label{sec:singsolve}

We want to solve the equation
\begin {equation}
  \left[ -\partial_{\tort}^2 - \tau_1 \, \frac{\kappa}{\tort^{2-\idx}} \right]
  \psi
  = \omega^2 \psi .
\end {equation}
It's convenient to think in terms of
eigenstates of $\tau_1$ and treat $\tau_1$ simply as a sign $\pm$
in what follows.
Accordingly,
defining the number $\bar\kappa \equiv \tau_1 \kappa$ and
dropping consideration of the spinor structure,
our equation has the form
\begin {equation}
  \left[ -\partial_{\tort}^2 - \frac{\bar\kappa}{\tort^{2-\idx}} \right]
  \phi
  = \omega^2 \phi .
\label {eq:Vfx}
\end {equation}
We don't know of a closed form solution to this equation.
However, in the large $\omega$ limit, we can see that it is
adequate to solve perturbatively in $\bar\kappa$.
Treating $\bar\kappa/\tort^{2-\idx}$ as small compared to $\omega^2$ requires
\begin {equation}
   |\tort| \gg \left|\frac{\bar\kappa}{\omega^2}\right|^{1/(2-\idx)} .
\label {eq:cond1}
\end {equation}
On the other hand, in order to match the WKB expansions
(\ref{eq:fWKBb}) to (\ref{eq:fWKBa}), we need the solution in
a region where neither $e^{+i \omega\tort}$ nor $e^{-i\omega\tort}$ becomes
exponentially small (and so small compared to the approximation error
in the other solution) at any time in the process of
rotating $\arg(\omega\tort)$ from $0$ to $\pi$.  That requires
\begin {equation}
   |\omega\tort| \lesssim 1 .
\label {eq:cond2}
\end {equation}
Fortunately, both conditions (\ref{eq:cond1}) and (\ref{eq:cond2})
can be simultaneously satisfied if
\begin {equation}
  |\bar\kappa| \ll |\omega|^\idx .
\label {eq:cond12}
\end {equation}
Since $\idx = (D-3)/2(D-2)$, (\ref{eq:cond12})
will always be satisfied for
sufficiently large $|\omega|$ at fixed $|\k|$ provided that
$D > 3$.%
\footnote{
  For the case $D=3$ and so $a=0$, one could solve (\ref{eq:Vfx}) directly
  in terms of Bessel functions, also incorporating the previously dropped
  first term
  in (\ref{eq:Vsing0}).  However, finding asymptotic WKB results is not
  important in the $D=3$ case because that's the case of a BTZ black
  hole, for which exact results are known \cite{BTZdirac}.
}
Using (\ref{eq:T}) and (\ref{eq:kappa}), the condition can be cast
into the form
\begin {equation}
  \frac{|\k|}{T} \ll \left(\frac{|\omega|}{T}\right)^{2-\idx} .
\end {equation}

The solutions to the unperturbed ($\bar\kappa{=}0$) version of
(\ref{eq:Vfx}) are simply $e^{\pm i\omega\tort}$.
So write $\phi = e^{\pm i\omega\tort} (1 + \bar\kappa\xi)$ and
linearize the equation in $\bar\kappa$, giving
\begin {equation}
  e^{\mp 2i\omega\tort}
  \partial_{\tort} \left( e^{\pm 2i\omega\tort} \partial_{\tort} \xi \right)
  \simeq - \frac{1}{\tort^{2-\idx}} \,.
\end {equation}
The corresponding solutions are
\begin {equation}
   \phi^{(\pm)} \simeq
   e^{\pm i\omega\tort}
   \left[
     1
     - \bar\kappa
       \int_{\tort}^\infty d\tort' \> e^{\mp 2i\omega\tort'}
       \int_{\tort'}^\infty d\tort'' \> \frac{e^{\pm 2i\omega\tort''}}{(\tort'')^{2-\idx}}
   \right] .
\end {equation}
The integral gives
\begin {equation}
   \phi^{(\pm)} \simeq
   e^{\pm i\omega\tort}
   + \bar\kappa \,
     \frac{\Gamma(\idx,\mp2i\omega\tort)}{(\mp2i\omega)^\idx (1-\idx)} \,
     e^{\mp i\omega\tort} ,
\label {eq:fsoln}
\end {equation}
where $\Gamma(a,z)$ is the incomplete $\Gamma$ function
\begin {equation}
   \Gamma(a,z)
   \equiv \int_z^\infty dt \> e^{-t} t^{a-1}
   \equiv \Gamma(a) - \gamma(a,z) .
\label {eq:Gamma}
\end {equation}


\subsubsection{The matching}

The asymptotic behavior of the incomplete $\Gamma$ function is
\begin {equation}
   \Gamma(a,z) \simeq z^{a-1} e^{-z}
   \qquad
   \mbox{($|z| \to \infty$ with $|\arg z| < \frac{3\pi}{2}$)} .
\label {eq:Gammax}
\end {equation}
For positive real $\omega\tort$,
the magnitude of the second term in the solution (\ref{eq:fsoln})
therefore falls algebraically for $\omega\tort \gg 1$ since
the exponent $a-1 = -(D-1)/2(D-2)$ is negative, and so
the asymptotic behavior of these solutions is
\begin {equation}
   \phi^{(\pm)} \to e^{\pm i \omega\tort}
   \qquad \mbox{($\omega\tort$ positive)} .
\label {eq:phi1}
\end {equation}

We now need to match this to the asymptotic behavior on the
negative $\omega\tort$ Stokes line in fig.\ \ref{fig:stokes}.
Near the origin,
moving from the positive $\omega\tort$ line clockwise to the negative
$\omega\tort$ line in that figure corresponds to
rotating $r$ by $e^{i\pi/(D-2)}$ and $\tort$ by $e^{i\pi}$.
Then, along the negative $\omega\tort$ Stokes line,
$\arg(-i \omega\tort) = \frac{\pi}{2}$ again satisfies the condition
for (\ref{eq:Gammax}), and so
\begin {equation}
   \phi^{(+)} \to e^{i \omega\tort}
   \qquad \mbox{($\omega \tort$ negative)}
\label {eq:phi2}
\end {equation}
just like in (\ref{eq:phi1}).

The interesting case is what happens to the asymptotic expansion
of $\phi^{(-)}$ when we analytically continue from the positive
$\omega\tort$ line to the negative $\omega\tort$ line.
In order to keep phases straight, it is convenient to rewrite
\begin {equation}
   \omega\tort = e^{i \pi} \omega y
\end {equation}
along the negative $\omega\tort$ Stokes line, with
$\omega y = -\omega\tort$ positive.
In this case, 
$\arg(i \omega\tort) = \frac{3\pi}{2}$, which does not satisfy
the condition of (\ref{eq:Gammax}).
From the integral formula
\begin {equation}
   \gamma(a,z)
   \equiv \int_0^z dt \> e^{-t} t^{a-1} ,
\label {eq:gamma}
\end {equation} 
one can show the monodromy relation
\begin {equation}
   \gamma(a, e^{i2\pi n}z) = e^{i 2\pi n a} \gamma(a,z)
\label {eq:gammaMonodromy}
\end {equation}
for integer $n$, and so
\begin {equation}
   \Gamma(a, e^{i2\pi n}z) = [1-e^{i2\pi na}] \Gamma(a) + e^{i 2\pi n a} \Gamma(a,z) .
\end {equation}
Using this relation, we can rewrite
\begin {equation}
   \Gamma(a,2i\omega\tort)
   = \Gamma(a, e^{i3\pi/2} 2 \omega y)
   = [1-e^{i2\pi a}] \Gamma(a) + e^{i 2\pi a} \, \Gamma(a,e^{-i\pi/2} 2 \omega y)
\end {equation}
and then use the standard expansion (\ref{eq:Gammax}) to show that the
last term disappears at large positive $\omega y$.  The result is that
(\ref{eq:fsoln}) yields
\begin {equation}
   \phi^{(-)} \to
   e^{-i \omega\tort}
   - \bar\kappa \,
     \frac{[1-e^{i2\pi\idx}]\,\Gamma(\idx)}{(2e^{i\pi/2}\omega)^\idx (1-\idx)} \,
     e^{i\omega\tort}
   \qquad \mbox{($\omega\tort$ negative)} .
\end {equation}
Using $\bar\kappa\equiv\tau_1\kappa$ and (\ref{eq:kappa}) for $\kappa$
and (\ref{eq:T}) for $T$, this can be written as
\begin {equation}
   \phi^{(-)} \to
   e^{-i \omega\tort}
   + \tau_1 \lambda \, 
     \frac{(k/\pi T)}{(\omega/\pi T)^\idx} \,
     e^{i\omega\tort}
   \qquad \mbox{($\omega\tort$ negative)} ,
\label {eq:phi3}
\end {equation}
where
\begin {equation}
   \lambda \equiv
   e^{-i\pi\idx/2} (1-\idx)^{1-\idx} \,
   \Gamma(\idx) \sin(\pi \idx) .
\label {eq:lambda}
\end {equation}

The upshot of (\ref{eq:phi1}), (\ref{eq:phi2}), and (\ref{eq:phi3})
is that an expansion
\begin {equation}
   \phi \simeq {\cal B}_+ e^{i\omega\tort} + {\cal B}_- e^{-i\omega\tort}
\end {equation}
on the positive $\omega\tort$ Stokes line corresponds to an expansion
\begin {equation}
   \phi \simeq
   \left[
      {\cal B}_+
      + \tau_1 \lambda \, 
        \frac{(k/\pi T)}{(\omega/\pi T)^\idx} \,
        {\cal B}_-
   \right] e^{i\omega\tort}
   + {\cal B}_- e^{-i\omega\tort}
\label {eq:fsingmatch}
\end {equation}
on the negative $\omega\tort$ Stokes line.
Writing the latter as
\begin {equation}
   \phi \simeq {\cal A}_+ e^{i\omega\tort} + {\cal A}_- e^{-i\omega\tort}
   \qquad \mbox{($\omega\tort$ negative)} ,
\label {eq:Aexpansion}
\end {equation}
the relationship is
\begin {equation}
  \begin{pmatrix} {\cal A}_+ \\ {\cal A}_- \end{pmatrix}
  =
  \begin{pmatrix}
    1 & 
    \tau_1 \lambda \, 
        \frac{(k/\pi T)}{(\omega/\pi T)^\idx} \\
    0 & 1
  \end{pmatrix}
  \begin{pmatrix} {\cal B}_+ \\ {\cal B}_- \end{pmatrix} .
\end {equation}


\subsection{Putting it all together}

We now impose the infalling boundary condition at the horizon, which is that
the $e^{i\omega\tort}$ term in the expansion (\ref{eq:Aexpansion}) must
vanish along the negative $\omega\tort$ Stokes line that connects
the singularity $r{=}0$ to the horizon.
From (\ref{eq:fsingmatch}) that condition is
\begin {equation}
   {\cal B}_+
   = - \tau_1 \lambda \,
        \frac{(k/\pi T)}{(\omega/\pi T)^\idx} \,
        {\cal B}_- .
\label {eq:fmatch1}
\end {equation}
On the other hand, along the positive $\omega\tort$ Stokes line,
comparison of the WKB form (\ref{eq:fWKBb}) to the
form (\ref{eq:boundarymatch})
that we got from applying the quasinormal mode condition at the
boundary ($r \to\infty$) gives
\begin {equation}
   {\cal B}_- = -i \tau_1 e^{2i\omega\torto-im\Rads\pi} {\cal B}_+ .
\label {eq:fmatch2}
\end {equation}
Consistency of (\ref{eq:fmatch1}) and (\ref{eq:fmatch2}) requires the
condition
\begin {equation}
   i\lambda\,\frac{(k/\pi T)}{(\omega/\pi T)^\idx}\,
   e^{2i\omega\torto-im\Rads\pi}
   = 1 ,
\end {equation}
which determines the quasi-normal mode frequencies.
This condition is satisfied when
\begin {equation}
   i \tfrac\pi2
   + \ln\left(\lambda\,\frac{(k/\pi T)}{(\omega/\pi T)^\idx}\right)
   + 2i\omega\torto-im\Rads\pi
   = i2\pi n
\end {equation}
for integer $n$.
Solving for $\omega$ by iteration in the large $n$ limit, and
using eq.\ (\ref{eq:x0}) for $\torto$ and (\ref{eq:lambda}) for $\lambda$,
then produces the result
\begin {multline}
   \frac{\omega_{n\pm}}{\pi T} \simeq
   4 e^{-i \pi/(D-1)} \sin\Bigl( \frac{\pi}{D-1} \Bigr)
\\ \times
   \left[
      n
      - \frac{i}{2\pi} \ln\left( \frac{n^\idx}{k/\pi T} \right)
      + \tfrac12(|m\Rads|-\tfrac12)
      + \frac{(D-3)^2}{8(D-1)(D-2)}
      + \frac{i \xi}{\pi}
   \right] ,
\label {eq:fermion2}
\end {multline}
with $\xi$ defined as in (\ref{eq:xi}).
Now recall from the definition (\ref{eq:k}) of $k$ that
$k = \pm |\k|$, depending on the spin state.
Eq.\ (\ref{eq:fermion2}) may then be recast in terms of $|\k|$ as
the result
(\ref{eq:fermion}) quoted in the introduction.%
\footnote{
  Note in the $k = -|\k|$ case
  that any $+ 2\pi i N$ ambiguity in the value of $\ln k = \ln(-|\k|)$
  may be absorbed through
  $O(n^0)$ by shifting the definition of $n$ in (\ref{eq:fermion})
  by an integer $N$, which is just a matter of labeling convention
  for the quasinormal modes.
}


\section {Numerical Method}
\label {sec:nmethod}

In this section, we discuss our numerical method for finding
quasinormal mode frequencies in the massive case,
which we will use to test the
asymptotic result (\ref{eq:fermion}).  We have not made exhaustive
comparisons to find the most efficient numerical algorithm in
the spin-$1/2$ case, but we will just use a variation on one of the
methods often used in the literature.
We will find a recursion relation for a series solution to the Dirac
equation expanded about the horizon.  We will use that recursion
relation to evaluate that series at the AdS
boundary as a function of $\omega$ (for given $m$ and $|\k|$).
Then we will search the complex $\omega$ plane to find values
$\omega{=}\omega_n$ where the value at the AdS boundary vanishes.



It is useful to have an equation that eliminates the spinor structure
from the Dirac equation (\ref{eq:Dirac}).  It is also convenient to have
an equation that does not involve square roots of $f$, since square
roots in our equation will not generate a recursion relation (for the
series solution) that has a bounded number of terms.
(However, if one must,
it is possible to get by with an unbounded number of terms in
the recursion relation, as in ref.\ \cite{GiammatteoJing}).

To obtain the equation we will use, start with the Dirac equation
(\ref{eq:Dirac}) and change basis to rewrite it in terms of
\begin {equation}
   A \equiv \frac{\psi_+ + i \psi_-}{\sqrt2} ,
   \qquad
   B \equiv \frac{i\psi_+ + \psi_-}{\sqrt2} ,
\end {equation}
to get
\begin {equation}
   \left[
      F \partial_r
      + i \omega \tau_3
      + m F^{1/2} \tau_2
      + f^{1/2} k \tau_1
   \right] 
   \begin{pmatrix} A \\ B \end{pmatrix}
   = 0 .
\end {equation}
In terms of components,
\begin {align}
   (F \partial_r + i \omega)A &=
   - \Bigl(k - \frac{imr}{\Rads}\Bigr) f^{1/2} B ,
\label {eq:A}
\\
   (F \partial_r - i \omega)B &=
   - \Bigl(k + \frac{imr}{\Rads}\Bigr) f^{1/2} A .
\end {align}
Rewrite the last as
\begin {equation}
   f^{-1/2}(F \partial_r - i \omega)f^{-1/2} f^{1/2} B =
   - \Bigl(k + \frac{imr}{\Rads}\Bigr) A .
\label{eq:B}
\end {equation}
Combining (\ref{eq:A}) and (\ref{eq:B}),
\begin {equation}
   f^{-1/2}(F \partial_r - i \omega)
   \left[ f^{-1/2} \Bigl(k-\frac{i mr}{\Rads}\Bigr)^{-1}
          (F \partial_r + i \omega) A \right]
   = \Bigl(k + \frac{imr}{\Rads}\Bigr) A ,
\end {equation}
which may be expanded to
\begin {equation}
   (F \partial_r)^2 A
   + \left(
        - \frac{r^2 \partial_r f}{2\Rads^2}
        + \frac{im F}{k\Rads-imr}
     \right)
     (F \partial_r + i \omega) A
   + (\omega^2 - m^2 F - k^2 f) A
   = 0 .
\label {eq:nice}
\end {equation}
This equation does not involve any square roots of $f$.  By multiplying
through by $k\Rads-imr$, it will be possible to write an equation whose
power-series solutions will have recursion relations with a fixed
number of terms.

We note in passing that the massless version of (\ref{eq:nice}) is
related to the Teukolsky equation which is often used to simultaneously
study massless fields of all different spins in $D{=}4$.  We
point out the translation to a selection of formulas in the literature
in appendix \ref{app:Teukolsky}.

Now factor out the behavior of the solution near the horizon and near
the boundary by writing
\begin {equation}
   A(z) = z^{-m\Rads} e^{-i\omega\tort} H(z) ,
\end {equation}
where $z = \Rads^2/r$.
The equation for $H$ is then
\begin {multline}
   z f \partial_z^2 H
   + \left( 2 i \omega z - 2 \bar m f + \frac{z \partial_z f}{2}
            - \frac{i \bar m f}{kz-i\bar m} \right)
     \partial_z H
\\
   - \left[
       \bar m^2 \, \frac{(1-f)}{z}
       + \bar m
         \left( 2 i \omega + \frac{\partial_z f}{2}
                - \frac{kf}{kz-i\bar m} \right)
       + k^2 z
     \right] H
   = 0 ,
\label {eq:Heq2}
\end {multline}
where we have introduced the dimensionless mass
\begin {equation}
   \bar m \equiv m \Rads .
\end {equation}
Multiplying through by $kz-i \bar m$,
and specializing now to units where $\zh = 1$,
one can obtain a
recursion relation for a series solution
\begin {equation}
   H(z) = \sum_{n=0}^\infty a_n (1-z)^n
\label {eq:Hseries}
\end {equation}
about the horizon.  For $D{=}5$, this is a 6-term recursion relation
for the coefficients $a_n$,
which is the shortest recursion relation we were able to find for
the massive case.
Specifically, for $D{=}5$, the choice of units $\zh=1$
is $\pi T = 1$ and the recursion relation is
\begin {equation}
   \sum_{j=0}^5 \alpha_{-j}(n) \, a_{n-j} = 0
\end {equation}
with
\begin {align}
   \alpha_{-5} &=
   -(4+\m-n)(5+\m-n)k ,
\\
   \alpha_{-4} &=
   (4+\m-n)\bigl[ (20+4\m-6n)k + (-2-\m+n)i\m \bigr] ,
\\
   \alpha_{-3} &=
   (-120-54\m-6\m^2+85n+20\m n-15n^2)k - k^3
\nonumber\\ & \qquad
   + (24+18\m+3\m^2-23n-8\m n+5n^2)i\m ,
\\
   \alpha_{-2} &=
   (80+36\m+4\m^2-80n-20\m n+20n^2)k + 2k^3
\nonumber\\ & \qquad
   + (-24-18\m-3\m^2+32n+12\m n-10n^2)i\m
   - i k^2\m + (4+2\m-2n)i k \omega ,
\\
   \alpha_{-1} &=
   (-18-6\m-\m^2+32n+8\m n-14n^2)k - k^3
\nonumber\\ & \qquad
   + (8+6\m+\m^2-18n-8\m n+10n^2)i\m
\nonumber\\ & \qquad
   + i k^2\m + (-2-2\m+2n)\m \omega + (-4-2\m+4n)i k \omega ,
\\
   \alpha_{0} &=
   2n(k-i\m)(-1+2n-i \omega) .
\end {align}

Our numerical method is to use this recursion relation to calculate
all the $a_n$ while computing the value of $H$ at the boundary as
\begin {equation}
   H(z{=}0) = \sum_{n=0}^\infty a_n
\label {eq:Hbdy}
\end {equation}
(cut off at some suitably high value of $n$), and then to scan
the complex $\omega$ plane to find zeros of (\ref{eq:Hbdy}).


\section {Numerics compared to asymptotic formula}
\label {sec:compare}

\subsection {Massive \boldmath$D{=}5$}

\begin {figure}
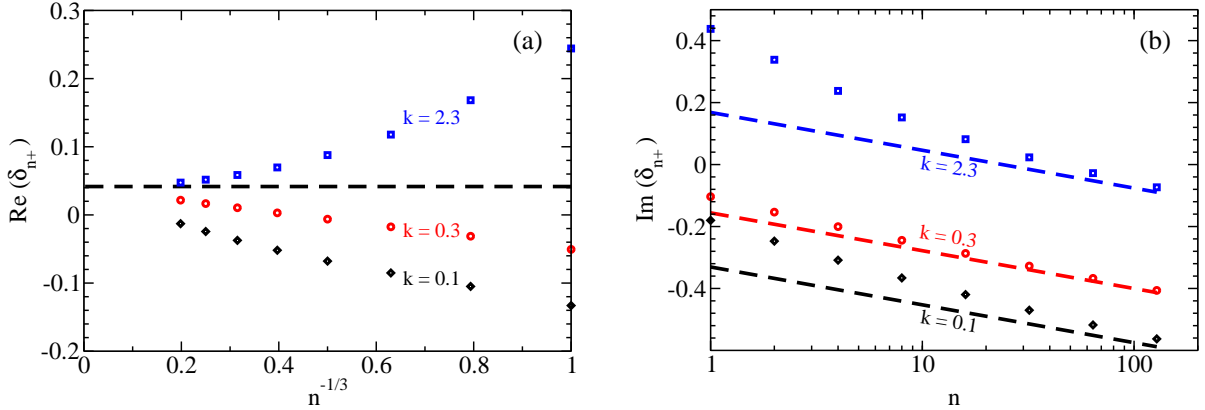

\begin {center}
  \includegraphics[scale=0.30]{tabre.eps}
  \hspace{0.2in}
  \includegraphics[scale=0.30]{tabim.eps}
  \caption{
     \label{fig:deltan}
     A plot of the (a) real and (b) imaginary parts of
     the offsets $\delta_n$ defined by (\ref{eq:deltan}) for
     $D=5$, $m\Rads=\tfrac12$, $|\k|/\pi T = 0.1$, $0.3$, and $2.3$,
     and the representative overtones $n=1,2,4,8,16,\cdots,128$.
     Data points are from our numerics, whereas the dashed lines
     indicate the value of $\delta_n$ that would be
     given by the asymptotic formula (\ref{eq:D5b}).
     Note the different choices made for the horizontal axis in the
     two figures, and that large $n$ corresponds to the left
     and right hand sides of (a) and (b) respectively.
  }
\end {center}
\end {figure}

We have already shown in fig.\ \ref{fig:wplane}
one comparison of our $D{=}5$
numerical results for $\omega_n$ with the
asymptotic formula (\ref{eq:D5})
In order to check more accurately, it is helpful to study much higher
$n$ and to plot the offset $\delta_n$ from the
leading $O(n)$ asymptotic formula, defined by
\begin {equation}
   \delta_n \equiv
   \frac{\omega_n}{\Delta\omega_\infty} - n
   \equiv
   \frac{\omega_n}{2(1-i)\pi T} - n
\label {eq:deltan}
\end {equation}
(for $D{=}5$).  Fig.\ \ref{fig:deltan} shows data points for
$\delta_{n+}$ from numerics%
\footnote{
  Some technical details: To automate the scan for zeros of
  (\ref{eq:Hbdy}), we started our search for each $n$ at the
  value $\omega=\omega_n-0.2 \,\m i$ with $\omega_n$
  given by the asymptotic formula
  (\ref{eq:D5b}).  Some readers may be surprised that our
  numerical method was accurate enough to reach overtones as
  high as $n=128$.  We achieved this by mindless brute force: we simply
  increased the precision of arithmetic used
  until we obtained stable results.
  For example, the $n=128$ results were computed using 3200-digit
  precision arithmetic.
}
plotted against dashed lines showing the
asymptotic result taken from (\ref{eq:D5}):
\begin {equation}
   \frac{\omega_{n\pm}}{\pi T} \simeq
   2(1-i)
   \left[
      n
      - \frac{i}{2\pi} \ln\left( \frac{n^{1/3}}{|\k|/\pi T} \right)
      + \tfrac12(|m\Rads|-1\pm\tfrac12)
      + \frac{1}{24}
      + \frac{i \xi_5}{\pi}
   \right] .
\label {eq:D5b}
\end {equation}
In particular, the
large $n$ behavior of $\Im\delta_{n+}$ in fig.\ \ref{fig:deltan}b clearly
shows agreement with the asymptotic $\log n$ and $|\k|$ dependence found in
(\ref{eq:D5b}).

For the sake of comparison for anyone in the future making similar calculations,
we provide tables of some of our numerical results in appendix
\ref{app:tables}.


\subsection {Massless \boldmath$D{=}4$}

\begin {figure}
\begin {center}
  \includegraphics[scale=0.60]{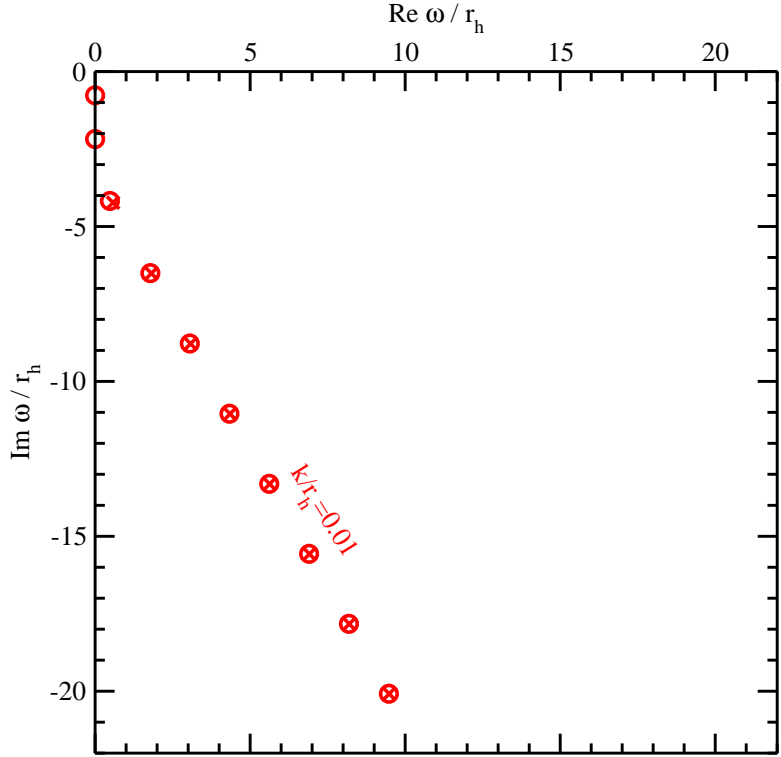}
  \caption{
     \label{fig:wplaneGJ}
     Retarded quasi-normal mode frequencies in the lower-right quadrant of the
     complex $\omega$ plane for massless $D{=}4$ Dirac fermions.
     The red circles are numerical results taken from the
     $\ell{=}0$ column of Table II of ref.\
     \cite{GiammatteoJing} and corresponds to $|\k| = 0.01 \, \rh$.
     The crosses show
     the asymptotic formula (\ref{eq:D4GJ}) for $\omega_{n+}$.
     (We do not also show the $\ell=1$ results here because they would
     crowd too close to the $\ell=0$ results.)
  }
\end {center}
\end {figure}

For the case of massless fermions in $D{=}4$,
fig.\ \ref{fig:wplaneGJ} shows a comparison of our
asymptotic formula (\ref{eq:fmassless4}) with
numerical results
of Giammatteo and Jing \cite{GiammatteoJing}.
In order to facilitate comparison, it is useful to use
(\ref{eq:T}) to recast
(\ref{eq:fmassless4}) as
\begin {equation}
   \frac{\omega\Rads^2}{\rh} \simeq
   e^{-i \pi/3} \frac{3^{3/2}}{2}
   \left[
      n
      - \frac{i}{2\pi} \ln\left( \frac{n^{1/4}}{4|\k|\Rads^2/3\rh} \right)
      + \frac{25}{48}
      + \frac{i \xi_4}{\pi}
   \right] .
\label {eq:D4GJ}
\end {equation}
Giammatteo and Jing study different sizes of black holes, whereas
our asymptotic formula was only derived in the large black hole
limit, and so we compare only to their results for their largest
black hole.
Their $r_1$ is our $\rh/\Rads$, and their largest black holes correspond
to
Table II of ref.\ \cite{GiammatteoJing} with $r_1=100$.
Their $k$ is an integer, related to the mode of spherical harmonics,
but translates in the large black hole limit into our $|\k|\Rads$.
In their Table II, the columns labeled $\ell=0$ and $\ell=1$
correspond to their $k=1$ and $k=2$ respectively.
Their dimensionless
$\omega$ quoted in numerical results is simply our $\omega\Rads$.
That is, they work in units where $\Rads=1$.

\begin {figure}
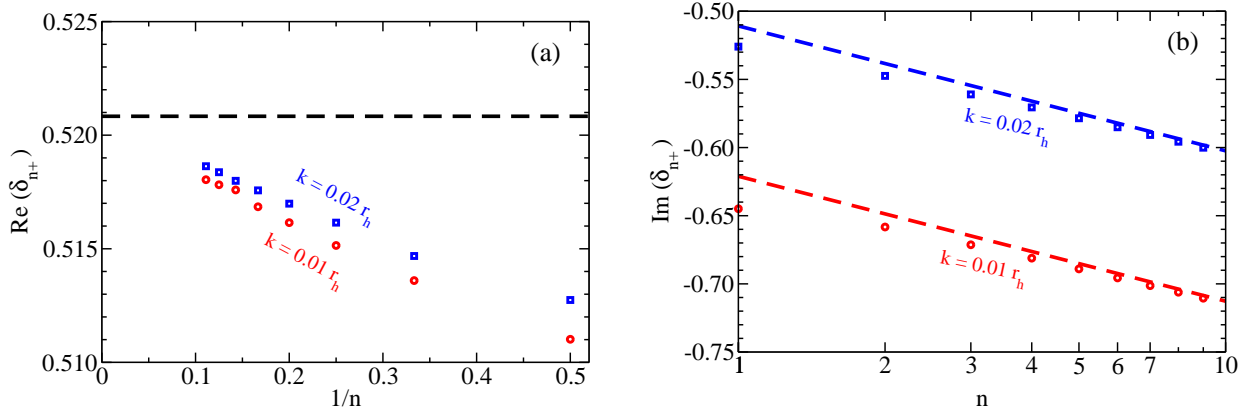

\begin {center}
  \includegraphics[scale=0.30]{tabreGJ.eps}
  \hspace{0.2in}
  \includegraphics[scale=0.30]{tabimGJ.eps}
  \caption{
     \label{fig:deltanGJ}
     The (a) real and (b) imaginary parts of
     the offsets $\delta_n$ defined by (\ref{eq:deltanGJ}) extracted
     from the $\ell=0$ and $\ell=1$ numerical results (corresponding
     to $|\k| = 0.01 \, \rh$ and $0.02 \, \rh$ respectively) of Table II of
     ref.\ \cite{GiammatteoJing} for massless Dirac fermions in
     $D{=}4$.
     The dashed lines
     indicate the value of $\delta_n$ that would be
     given by the asymptotic formula (\ref{eq:D4GJ}).
     In order to align the overtones found in
     ref.\ \cite{GiammatteoJing} with the ones given by our asymptotic
     formula, the overtone number $n$ we use in this figure corresponds
     to their $n-1$.
  }
\end {center}
\end {figure}

As in the earlier discussion of $D{=}5$, we also look at the offset
$\delta_n$, which in the $D{=}4$ case is
\begin {equation}
   \delta_n \equiv
   \frac{\omega_n}{\Delta\omega_\infty} - n
   \equiv
   \frac{\omega_n}{(\frac{3\sqrt 3}{4}-\frac{9i}{4})\rh} - n .
\label {eq:deltanGJ}
\end {equation}
Fig.\ \ref{fig:deltanGJ} shows data points for
$\delta_{n+}$ from Giammatteo and Jing plotted against dashed lines
showing the
asymptotic result taken from (\ref{eq:D4GJ}).
The agreement at large $n$ (and even small $n$) is quite good.

\begin{acknowledgments}

We thank Diana Vaman for many helpful conversations.
This work was supported, in part, by the U.S. Department
of Energy under Grant No.~DE-SC0007984.

\end{acknowledgments}


\appendix

\section {Relation to Teukolsky equation}
\label {app:Teukolsky}

In this appendix, it will be convenient to set $L{=}1$.
In the massless case, the form (\ref{eq:nice}) of the Dirac equation
becomes
\begin {equation}
   (F \partial_r)^2 A
   - \frac{r^2 \partial_r f}{2}
     (F \partial_r + i \omega) A
   + (\omega^2 - k^2 f) A
   = 0 ,
\end {equation}
If one now defines $\psi_{1/2}$ by
\begin {equation}
   A = f^{1/4} \psi_{1/2},
\end {equation}
then the equation becomes
the Schr\"odinger-like problem
\begin {subequations}
\label {eq:T1}
\begin {equation}
   \left[ - \partial_{\tort}^2 + V_{1/2} \right] \psi_{1/2} = \omega^2 \psi_{1/2}
\end {equation}
with
\begin {equation}
   V_{1/2}(r) =
   -\tfrac 14 \, F \partial_r(r^2 f')
   +\tfrac1{16} \, r^4 (f')^2
   + \tfrac12 \, i \omega r^2 f'
   + k^2 f .
\end {equation}
\end {subequations}

Though the above equations are valid in any dimension, they agree
with the spin-$\tfrac12$ case of the Teukolsky equation
\cite{Teukolsky}.
The Teukolsky equation gives a unified description of arbitrary-spin
massless fields in certain $D{=}4$ metrics, and has been used
by a number of authors to study quasinormal modes for asymptotically
flat Schwarzschild and Kerr-Newman.
Our eqs.\ (\ref{eq:T1}) are equivalent, for example, to the particular
form of the Teukolsky equation given by Jing \cite{Jing}.%
\footnote{
   Specifically, see eqs. (2.16--17) of ref.\ \cite{Jing},
   taking $s=\tfrac12$, and identifying Jing's $\Delta$ with
   our $r^2 F = r^4 f$ and
   Jing's $\lambda^2 \equiv \bigl(\ell+\tfrac12\bigr)^2$
   with our $k^2$.  In earlier discussion, Jing's
   $\mathds{R}_{1/2}$ and $\mathds{R}_{-1/2}$ are our $A/rF^{1/2}$
   and $-B$.
}
For more specific comparison to papers on asymptotically flat
$D{=}4$ Schwarzschild black holes, note that
the metric (\ref{eq:metricr}) has the same form as
an asymptotically flat $D{=}4$ black hole metric if one identifies
\begin {equation}
   F = 1 - \frac{\rh}{r} \,,
   \qquad
   f \equiv \frac{F}{r^2} \,,
\end {equation}
and takes $\x$ to be angular variables approximated as flat
(an approximation that will make sense in the limit of large
spherical harmonics).

For discussion of the Teukolsky equation in asymptotically AdS$_4$
spacetime, see ref.\ \cite{GiammatteoMoss}.
We our unaware of any generalization of the Teukolsky equation for
arbitrary spin to $D>4$.


\section {Tabulated numerical results for exact quasinormal frequencies}
\label {app:tables}

For the sake of anyone who might want to someday compare their own numerics
to our results, we tabulate some quasinormal mode frequencies in table
\ref{tab:QNM}.



\newpage

\begin {table}
\begin {tabular}{|c|ll|ll|ll|}
\hline
  $n\pm$
  & \multicolumn{6}{|c|}{$\omega_{n\pm}/\pi T$}
\\ \cline{2-7}
  & \multicolumn{2}{|c|}{$|\k| = 0.1 \, \pi T$}
  & \multicolumn{2}{|c|}{$|\k| = 0.3 \, \pi T$}
  & \multicolumn{2}{|c|}{$|\k| = 2.3 \, \pi T$}
\\
\hline
1$-$&1.09696&-1.80921 $i$&1.10987&-1.49754 $i$&2.75050&-0.76673 $i$\\
1$+$&1.37370&-2.09426 $i$&1.69167&-2.10603 $i$&3.36431&-1.61242 $i$\\
2$-$&2.91046&-3.98067 $i$&2.91651&-3.60052 $i$&4.18969&-2.70882 $i$\\
2$+$&3.29582&-4.28446 $i$&3.63039&-4.24441 $i$&5.01324&-3.65993 $i$\\
3$-$&4.80706&-6.05121 $i$&4.83157&-5.64176 $i$&5.93557&-4.75183 $i$\\
3$+$&5.24680&-6.38134 $i$&5.59180&-6.31677 $i$&6.82779&-5.71511 $i$\\
4$-$&6.73512&-8.09412 $i$&6.77754&-7.66773 $i$&7.78860&-6.79412 $i$\\
4$+$&7.21246&-8.44691 $i$&7.56440&-8.36564 $i$&8.71124&-7.76006 $i$\\
5$-$&8.68030&-10.12411 $i$&8.73829&-9.68687 $i$&9.69036&-8.82896 $i$\\
5$+$&9.18625&-10.49627 $i$&9.54318&-10.40232 $i$&10.62949&-9.79610 $i$\\
6$-$&10.63622&-12.14674 $i$&10.70765&-11.70209 $i$&11.61867&-10.85756 $i$\\
6$+$&11.16516&-12.53573 $i$&11.52585&-12.43157 $i$&12.56796&-11.82560 $i$\\
7$-$&12.59951&-14.16467 $i$&12.68262&-13.71478 $i$&13.56322&-12.88148 $i$\\
7$+$&13.14755&-14.56851 $i$&13.51120&-14.45582 $i$&14.51937&-13.85031 $i$\\
8$-$&14.56816&-16.17939 $i$&14.66151&-15.72568 $i$&15.51857&-14.90188 $i$\\
8$+$&15.13247&-16.59648 $i$&15.49851&-16.47650 $i$&16.47963&-15.87145 $i$\\
9$-$&16.54088&-18.19178 $i$&16.64330&-17.73526 $i$&17.48153&-16.91957 $i$\\
9$+$&17.11930&-18.62085 $i$&17.48731&-18.49450 $i$&18.44628&-17.88985 $i$\\
10$-$&18.51678&-20.20244 $i$&18.62730&-19.74380 $i$&19.45007&-18.93514 $i$\\
10$+$&19.10763&-20.64240 $i$&19.47728&-20.51042 $i$&20.41770&-19.90608 $i$\\
11$-$&20.49524&-22.21175 $i$&20.61306&-21.75153 $i$&21.42288&-20.94902 $i$\\
11$+$&21.09715&-22.66170 $i$&21.46821&-22.52467 $i$&22.39281&-21.92057 $i$\\
12$-$&22.47579&-24.22000 $i$&22.60023&-23.75858 $i$&23.39904&-22.96152 $i$\\
12$+$&23.08766&-24.67915 $i$&23.45992&-24.53757 $i$&24.37085&-23.93364 $i$\\
13$-$&24.45809&-26.22738 $i$&24.58857&-25.76508 $i$&25.37786&-24.97286 $i$\\
13$+$&25.07898&-26.69508 $i$&25.45229&-26.54933 $i$&26.35126&-25.94552 $i$\\
14$-$&26.44186&-28.23405 $i$&26.57789&-27.77110 $i$&27.35888&-26.98325 $i$\\
14$+$&27.07099&-28.70971 $i$&27.44522&-28.56015 $i$&28.33362&-27.95641 $i$\\
15$-$&28.42690&-30.24012 $i$&28.56804&-29.77672 $i$&29.34171&-28.99281 $i$\\
15$+$&29.06360&-30.72324 $i$&29.43864&-30.57015 $i$&30.31760&-29.96644 $i$\\
16$-$&30.41304&-32.24569 $i$&30.55890&-31.78198 $i$&31.32606&-31.00166 $i$\\
16$+$&31.05672&-32.73580 $i$&31.43247&-32.57945 $i$&32.30296&-31.97574 $i$\\
\hline
32$-$&62.27210&-64.30027 $i$&62.46486&-63.83934 $i$&63.17895&-63.09203 $i$\\
32$+$&62.98480&-64.86574 $i$&63.36594&-64.67616 $i$&64.16344&-64.07078 $i$\\
\hline
64$-$&126.1455&-128.3494 $i$&126.3765&-127.8987 $i$&127.0593&-127.1758 $i$\\
64$+$&126.9158&-128.9869 $i$&127.2987&-128.7680 $i$&128.0481&-128.1587 $i$\\
\hline
128$-$&254.0314&-256.3974 $i$&254.2921&-255.9603 $i$&254.9567&-255.2553 $i$\\
128$+$&254.8486&-257.0999 $i$&255.2307&-256.8559 $i$&255.9482&-256.2417 $i$\\
\hline
\end {tabular}
\caption
    {
    \label {tab:QNM}
    Numerical results for $\omega_{n\pm}$ in right-hand complex plane
    for $D{=}5$ and $m\Rads=\tfrac12$ .
    }
\end {table}



\begin{thebibliography}{}

\bibitem{QNMreview1}
  E.~Berti, V.~Cardoso and A.~O.~Starinets,
  ``Quasinormal modes of black holes and black branes,''
  Class.\ Quant.\ Grav.\  {\bf 26}, 163001 (2009)
  [arXiv:0905.2975 [gr-qc]].

\bibitem{QNMreview2}
  R.~A.~Konoplya and A.~Zhidenko,
  ``Quasinormal modes of black holes: From astrophysics to string theory,''
  Rev.\ Mod.\ Phys.\  {\bf 83}, 793 (2011)
  [arXiv:1102.4014 [gr-qc]].


\bibitem{Birmingham} 
  D.~Birmingham, I.~Sachs and S.~N.~Solodukhin,
  ``Conformal field theory interpretation of black hole quasinormal modes,''
  Phys.\ Rev.\ Lett.\  {\bf 88}, 151301 (2002)
  [hep-th/0112055].

\bibitem{SonStarinets} 
  D.~T.~Son and A.~O.~Starinets,
  ``Minkowski space correlators in AdS / CFT correspondence: Recipe and applications,''
  JHEP {\bf 0209}, 042 (2002)
  [hep-th/0205051].

\bibitem{Starinets} 
  A.~O.~Starinets,
  ``Quasinormal modes of near extremal black branes,''
  Phys.\ Rev.\ D {\bf 66}, 124013 (2002)
  [hep-th/0207133].

\bibitem{KovtunStarinets} 
  P.~K.~Kovtun and A.~O.~Starinets,
  ``Quasinormal modes and holography,''
  Phys.\ Rev.\ D {\bf 72}, 086009 (2005)
  [hep-th/0506184].

\bibitem{DHS}
  F.~Denef, S.~A.~Hartnoll and S.~Sachdev,
  ``Black hole determinants and quasinormal modes,''
  Class.\ Quant.\ Grav.\  {\bf 27}, 125001 (2010)
  [arXiv:0908.2657 [hep-th]].

\bibitem{DHS0}
  F.~Denef, S.~A.~Hartnoll and S.~Sachdev,
  ``Quantum oscillations and black hole ringing,''
  Phys.\ Rev.\ D {\bf 80}, 126016 (2009)
  [arXiv:0908.1788 [hep-th]].

\bibitem{BTZblackhole}
  M.~Banados, C.~Teitelboim and J.~Zanelli,
  ``The Black hole in three-dimensional space-time,''
  Phys.\ Rev.\ Lett.\  {\bf 69}, 1849 (1992)
  [hep-th/9204099].

\bibitem{BTZdirac}
  S.~Datta and J.~R.~David,
  ``Higher spin fermions in the BTZ black hole,''
  JHEP {\bf 1207}, 079 (2012)
  [arXiv:1202.5831 [hep-th]].

\bibitem{MusiriSiopsis:Dirac}
  S.~Musiri and G.~Siopsis,
  ``Perturbative calculation of quasi-normal modes of arbitrary spin
    in Schwarzschild spacetime,''
  Phys.\ Lett.\ B {\bf 650}, 279 (2007).

\bibitem{NatarioSchiappa}
  J.~Natario and R.~Schiappa,
  ``On the classification of asymptotic quasinormal frequencies for
    $d$-dimensional black holes and quantum gravity,''
  Adv.\ Theor.\ Math.\ Phys.\  {\bf 8}, 1001 (2004)
  [hep-th/0411267].

\bibitem{CardosoNatarioSchiappa}
  V.~Cardoso, J.~Natario and R.~Schiappa,
  ``Asymptotic quasinormal frequencies for black holes
    in nonasymptotically flat space-times,''
  J.\ Math.\ Phys.\  {\bf 45}, 4698 (2004)
  [hep-th/0403132].

\bibitem{MusiriNessSiopsis}
  S.~Musiri, S.~Ness and G.~Siopsis,
  ``Perturbative calculation of quasi-normal modes of
    AdS Schwarzschild black holes,''
  Phys.\ Rev.\ D {\bf 73}, 064001 (2006)
  [hep-th/0511113].

\bibitem{GiammatteoJing}
  M.~Giammatteo and J.~Jing,
  ``Dirac quasinormal frequencies in Schwarzschild-AdS space-time,''
  Phys.\ Rev.\ D {\bf 71}, 024007 (2005)
  [gr-qc/0403030].

\bibitem{KRvN}
  H.~J.~Kim, L.~J.~Romans and P.~van Nieuwenhuizen,
  ``Mass spectrum of chiral ten-dimensional $N{=}2$ Supergravity on $S^5$,''
  Phys.\ Rev.\ D {\bf 32}, 389 (1985).

\bibitem{HerzogRen}
  C.~P.~Herzog and J.~Ren,
  ``The Spin of Holographic Electrons at Nonzero Density and Temperature,''
  JHEP {\bf 1206}, 078 (2012)
  [arXiv:1204.0518 [hep-th]].

\bibitem{largem1}
  L.~Fidkowski, V.~Hubeny, M.~Kleban and S.~Shenker,
  ``The Black hole singularity in AdS / CFT,''
  JHEP {\bf 0402}, 014 (2004)
  [hep-th/0306170].

\bibitem{largem2}
  G.~Siopsis,
  ``Large mass expansion of quasinormal modes in AdS(5),''
  Phys.\ Lett.\ B {\bf 590}, 105 (2004)
  [hep-th/0402083].

\bibitem{largek}
  G.~Festuccia and H.~Liu,
  ``A Bohr-Sommerfeld quantization formula for quasinormal
    frequencies of AdS black holes,''
  Adv.\ Sci.\ Lett.\  {\bf 2}, 221 (2009)
  [arXiv:0811.1033 [gr-qc]].

\bibitem{MAGOO}
  O.~Aharony, S.~S.~Gubser, J.~M.~Maldacena, H.~Ooguri and Y.~Oz,
  ``Large N field theories, string theory and gravity,''
  Phys.\ Rept.\  {\bf 323}, 183 (2000)
  [hep-th/9905111].

\bibitem{MusiriSiopsis:AdS}
  S.~Musiri and G.~Siopsis,
  ``Asymptotic form of quasinormal modes of large AdS black holes,''
  Phys.\ Lett.\ B {\bf 576}, 309 (2003)
  [hep-th/0308196].

\bibitem{Giecold}
  G.~C.~Giecold,
  ``Fermionic Schwinger-Keldysh Propagators from AdS/CFT,''
  JHEP {\bf 0910}, 057 (2009)
  [arXiv:0904.4869 [hep-th]].

\bibitem{Henningson} 
  M.~Henningson and K.~Sfetsos,
  ``Spinors and the AdS / CFT correspondence,''
  Phys.\ Lett.\ B {\bf 431}, 63 (1998)
  [hep-th/9803251].

\bibitem{Leigh} 
  R.~G.~Leigh and M.~Rozali,
  ``The Large N limit of the (2,0) superconformal field theory,''
  Phys.\ Lett.\ B {\bf 431}, 311 (1998)
  [hep-th/9803068].

\bibitem{Teukolsky}
  S.~A.~Teukolsky,
  ``Rotating black holes: separable wave equations for gravitational
    and electromagnetic perturbations,''
  Phys.\ Rev.\ Lett.\  {\bf 29}, 1114 (1972);
  ``Perturbations of a rotating black hole. I. Fundamental equations
    for gravitational electromagnetic and neutrino field perturbations,''
  Astrophys.\ J.\  {\bf 185}, 635 (1973).


\bibitem{Jing}
  J.~-l.~Jing,
  ``Dirac quasinormal modes of Schwarzschild black hole,''
  Phys.\ Rev.\ D {\bf 71}, 124006 (2005)
  [gr-qc/0502023].

\bibitem {GiammatteoMoss}
  M.~Giammatteo and I.~G.~Moss,
  ``Gravitational quasinormal modes for Kerr anti-de Sitter black holes,''
  Class.\ Quant.\ Grav.\  {\bf 22}, 1803 (2005)
  [gr-qc/0502046].

\end{thebibliography}
\end {document}